# Data Analysis Methods Preliminaries for a Photon-based Hardware Random Number Generator

*Dmitriy* Beznosko[1*]*, Keith* Driscoll[2]*, Fernando* Guadarrama[3]*, Steven* Mai[3], and *Nikolas* Thornton[3]

[1] College of STEM, School of Sciences, Clayton State University, Morrow, GA 30260 USA
[2] College of STEM, Department of Mathematics, Clayton State University, Morrow, GA 30260 USA
[3] College of STEM, CSIT Department, Clayton State University, Morrow, GA 30260 USA

**Abstract.** Hardware random number generators (HRNG) are extensively used in the computer world for security purposes as well as in the science world as a source of high-quality randomness for the models and simulations. Currently existing HRNG are either costly or slow and of questionable quality. This work proposes a simple design of the HRNG based on the low-number photon absorption by a detector (a photo-multiplier tube of a silicon-based photodetector) that can provide a large volume of high-quality random numbers [1]. The different options of processing and the testing of quality of the generator output are described.

## 1 Introduction

High quality random numbers are necessary in the modern world. Ranging from encryption keys in cyber security to models and simulations for scientific use: it's important that these random numbers are of high quality and quickly attainable.

One common solution to the generation of random numbers is that of *pseudo*-random number generators, or PRNGs. PRNGs generate random numbers by first quantifying some unpredictable phenomena into a number or string and feeding it into an algorithm which yields numbers randomly based on that "seed". Easy places to find seeds include the user's mouse movements or the machine's uptime. These are only pseudorandom, however, as if given the same seed twice, the PRNG would generate the same "random" output. This is great for games like Minecraft, but not so great for cybersecurity encryption key generation.

By using a hardware random number generator (HRNG), random numbers that are not susceptible to the flaws found in PRNGs can be attained at a high rate.

### 1.1 Photon-based Solution

By using photon-detectors, which detect photons of varying amplitudes and multitudes due to quantum phenomena, high quality random numbers can be harvested. Two methods of detecting photons are explored in this article, one using a photo-multiplier tube (PMT) and the other using a multi-pixel photon counter (MPPC) from Hamamatsu [2]. The advantages of either setup as well as the quality of their output are explored.

## 2 Experimental Setup

For the PMT setup, a Hamamatsu H11284-30 (former R7723) PMT is placed into a blackout-box alongside a light emitting diode (LED), as shown in Figure 1. The PMT is biased at 1750 V and is connected to CAEN [3] DT5730 Flash Analog-to-Digital Converter (FADC). The LED is powered but the pulse that is about 50ns wide and 3V in amplitude. The PMT and ADC choice is based on the availability, it has been provided by the DUCK experiment [4] for this test. Note that the physical basis of the test is independent of and will not be affected by different hardware models. Further information on the hardware details is available in [5].

---

* Corresponding author: dmitriybeznosko@clayton.edu

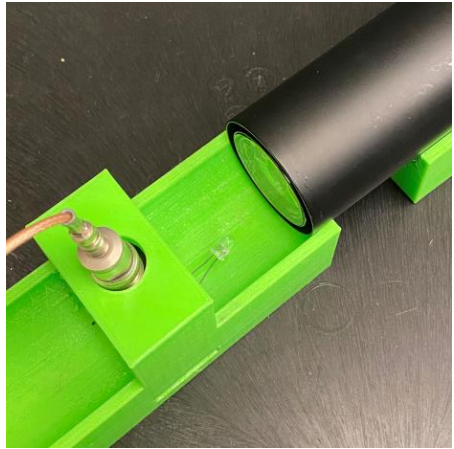

**Fig. 1.** PMT HRNG Experimental Setup. A 3D printed holder houses the PMT and LED in the dark box (not shown).

The MPPC setup is shown in Figure 2. The 1.3x1.3 mm$^2$ MPPC is connected to an Arduino Nano that also controls the attached LED as well as the Hamamatsu C11204-01 power supply providing 70V bias to the MPPC. The MPPC's output is connected to the Arduino ADC, which reads amplitudes in sync with an LED pulse. In the current setup, the LED is voltage-regulated using Arduino's default Pulse Width Modulation solution to low-voltage output. The Arduino communicates the amplitudes via serial data to the connected computer.

Note that the setups have been constructed and data is being collected. In section 3 we discuss the possible analysis methods.

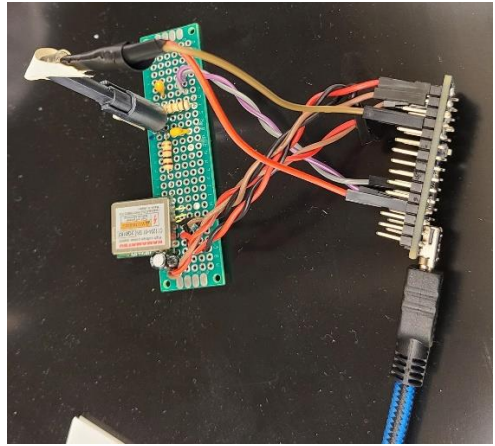

**Fig. 2.** MPPC HRNG Experimental Setup. MPPC sensor is in the light-tight housing and can operate under ambient light.

## 3 Data Analysis

Both setups output amplitude data as a text file that requires further processing to generate a random bitstream. Two methods of processing have been implemented.

The first processing method is that of High/Low. This method looks at every amplitude and compares if it is higher or lower than the previous amplitude. While obviously flawed, it was initially implemented as a testing code, and later was used to generate low-quality data for comparison purposes. If it is higher, then it adds a 1 to the bitstream. If it is lower, then it adds a 0. This is described in pseudo-code as shown in Figure 3.

```
if amp[i] > amp[i-1]: add 1 to bitstream
if amp[i] < amp[i-1]: add 0 to bitstream
else: skip
```

**Fig. 3.** High/low processing method pseudo-code

The second solution that is used for main data processing is the Even/Odd method. This method looks at every amplitude and determines if it is even or odd noting that Arduino board reports amplitudes as integers. If the amplitude value is even, then 1 is added to the bitstream. If the value is odd, then 0 is added. This is described in pseudo-code as shown in Figure 4.

```
if amp[i] % 2 == 0: add 1 to bitstream
else: add 0 bitstream
```

**Fig. 4.** Even/odd processing method pseudo-code

## 4 Testing Methods

For this work, 3 tests have been selected and implemented for determining the quality of output from both processing methods applied to each HRNG setup data. These tests are as follows: the Arithmetic Mean and Standard Deviation (AMSD) test, the Monte Carlo Pi Estimation (MCPE) test, and the Fractional Line Symmetry (FLS) test. The FLS test is a new test for randomness of the data and has been developed specifically for this project.

### 4.1 AMSD Test

Taking the mean and standard deviation of the bitstream is the simplest of the tests that are executed on all the bitstreams. It is effective at determining if there is more of one type of bit over the other but fails in detection of patterns.

An ideal result from this test would yield a near-perfect 0.5 for the mean of the bitstream and a near-perfect 0.5 for the standard deviation for the bitstreams of sufficient size.

### 4.2 MCPE Test

The MCPE Test is done by assigning two random numbers as x and y coordinates for within a square and counting each time a point falls within the quarter-cross-section of a circle ($x^2 + y^2 \leq radius$), and then dividing that total by the total number of points everywhere, $\pi$ can be estimated as $4 \cdot Red\ dots/All\ dots$, as shown in Figure 5 (left). By turning every 8 bits into a byte, turning every byte into a float between 1 and 0, and then pairing every byte to form x, y coordinate, points are processed as shown in the pseudo code in Figure 5 (right).

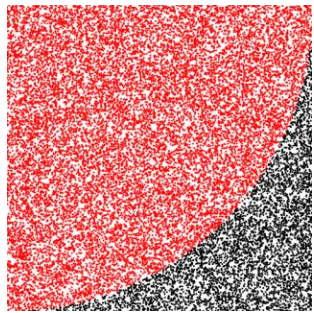

```
all_dots += 1
if X^2 + Y^2 <=1: red_dots += 1
pi = 4*(red_dots/all_dots)
```

**Fig. 5.** A square with red marking dots that fall within circle (left). MCPE pseudo-code (right).

An ideal result from this test would yield a value of $\pi$ accurate to several decimal places, depending on the size of the bitstream and therefore the number of coordinates. To account for the testing of bitstreams of various sizes, Python's Random library is put to the test using MCPE with the same number of coordinates generated as in the bitstream to see if any deviation could just be due to an insufficient coordinate count.

This test is very sensitive to the quality of randomness of the input data. When tested with simple PRNGs, above certain amount of the random numbers generated that goes over the capabilities of the library the value of $\pi$ may deviate in different ways, sometimes it starts oscillating around some value, sometimes growing up to 3.2 and above or reducing to 3 and lower.

### 4.3 FLS Test

The Fractional Line Symmetry Test (FLS Test) is a test developed specifically for this project that compares how frequently bits appear back-to-back horizontally and vertically when stacked and visualized into a 2D image. The folding done to stack and visualize the data is inherently random to the bitstream length itself. Once the linear bitstream is turned into a 2D image, the number of "lines" (back-to-back bits of some length) found horizontally and vertically should be the same.

This test is sensitive to poor quality random numbers that otherwise pass tests like the standard deviation and average of the bitstream. The number of lines that should be found by this test for any size bitstream is estimated using equation 1.

$$Lines\# = \frac{n+1}{(2^{n+1})n} \times bitstream\ length \quad (1)$$

where n is the line search length and the bitstream length is the number of total bits in the bitstream.

### 4.3.1 Visualization

Visualization is done by first assigning the bits colors. 1s are white pixels and 0s are black pixels. Figure 7 shows an example for how the bits are initially visualized. The bitstream used in Figure 6 will be used for the rest of this section as an example.

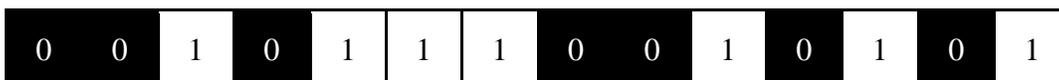

**Fig. 6.** Bitstream example

To ensure that the bits will eventually be able to be stacked into a perfect square to form the 2D image, the length of the bitstream must be increased. These "nothing bits" that are added are not actually a part of the bitstream, and only used for the sake of visualization in the FLS Test. To achieve this, the ceiling of the square root of the bitstream length must be taken and squared as shown in equation 2.

$$ceil(\sqrt{14})^2 = 16 \quad (2)$$

This means that $16 - 14 = 2$ new "nothing bits" must be added to the bitstream, which for the sake of visualization are shown as grey pixels, as seen in Figure 7. Now that the bitstream length is a square number, it can be stacked left to right into a perfect square, as shown in Figure 8.

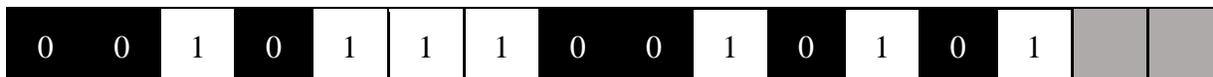

**Fig. 7.** Bits visualized with "nothing bits" added to the end.

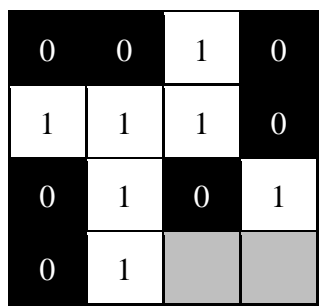

**Fig. 8.** The bitstream stacked into a square image.

### 4.3.2 Line Counting

Line counting is the act of counting how many times a 1 or a 0 appears back-to-back a certain number of times in a row. How many times they must appear back-to-back before being considered one full line is called the detect length. It's possible for bitstreams to have fractional lines. If a full line is found that has additional bits added towards the end, the additional bits are added to the total line count as $\frac{1}{detect\ length}$. Using a different bitstream than what has been used in previous figures, Figure 9 illustrates line searching for zeros with a detect length of 2.

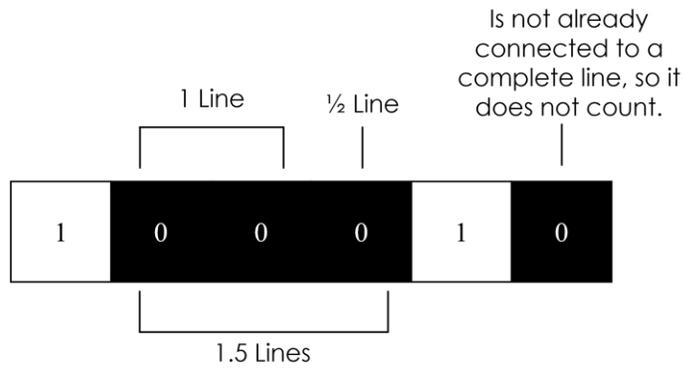

**Fig. 9.** How fractional lines are counted.

### 4.3.3 Horizontal Line Counting

Horizontal line counting is easier than vertical line counting. For horizontal lines, the bitstream does not need to be stacked at any point. It can be read left-to-right linearly and searched for lines as described in 4.3.2. Figure 10 shows the bitstream used in Figure 8 when line searched for zeros with a detect length of 2. The lines found are highlighted in blue. Figure 11 shows this bitstream when stacked and visualized, which demonstrates how the lines can cross the image-borders.

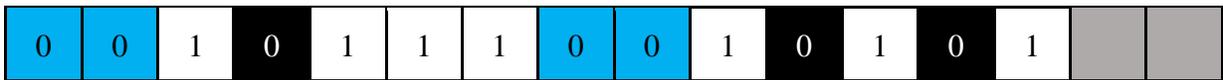

**Fig. 10.** Horizontal line counting for zeros with a detect length of two.

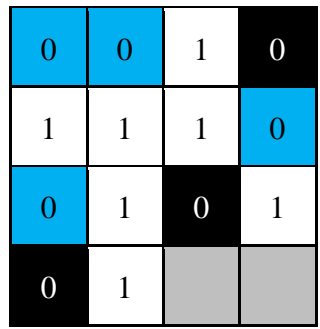

**Fig. 11.** Horizontal line counting when stacked and visualized.

### 4.3.4 Vertical Line Counting

Vertical line counting is more difficult than horizontal line counting, as it requires that the bits are first visualized and stacked as shown previously in Figure 8. After the bits are visualized the resulting 2D image is rotated 90 degrees counterclockwise as shown in Figure 12. From this, the bits are read left-to-right starting at the top left and unstacked into a linear bitstream. This is just the reverse of the visualization process described in 4.3.1. Figure 13 shows the bitstream after it has been unstacked.

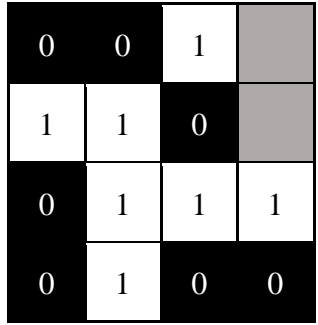

**Fig. 12.** Visualized and stacked bitstream after being rotated 90 degrees counterclockwise.

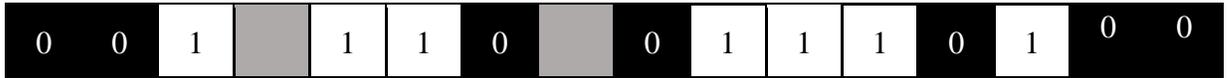

**Fig. 13.** The bitstream after being unstacked.

Now that the bitstream is linear, it can easily be searched for lines using the same method as horizontal lines. Figure 14 shows the bitstream after it has been line searched for zeros with a detect length of 2. The lines found are highlighted in red. Note that a grey "nothing bit" exists in between a line, but it does not stop it from being counted as a line. This bitstream can now be restacked into a square image, and then rotated back 90 degrees clockwise, as shown in Figure 15.

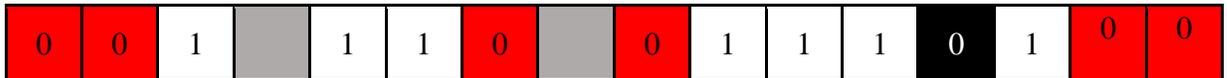

**Fig. 14.** Unstacked bitstream after being searched for lines.

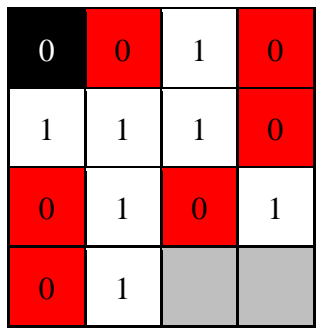

**Fig. 15.** Vertical line counting when stacked and visualized.

From this visualization, vertical line counting effectively counts lines starting at the top-right of the image and reading downward.

*4.3.5 Estimating Line Counts Empirical*

It is possible to calculate the probability of how many lines should appear in a bitstream by using Equation 1. This possibility was empirically found by observing patterns in a bitstream of $10^7$ random bits generated using Python's Random library. Figure 16 shows the number of lines detected for each detect length in these $10^7$ random bits on a base-2 logarithmic scale. By taking the lines found for each detect length, dividing them by the total number of bits in the bitstream, and multiplying them by the detect length, a series labeled $L_n$ is found as shown in Table 1.

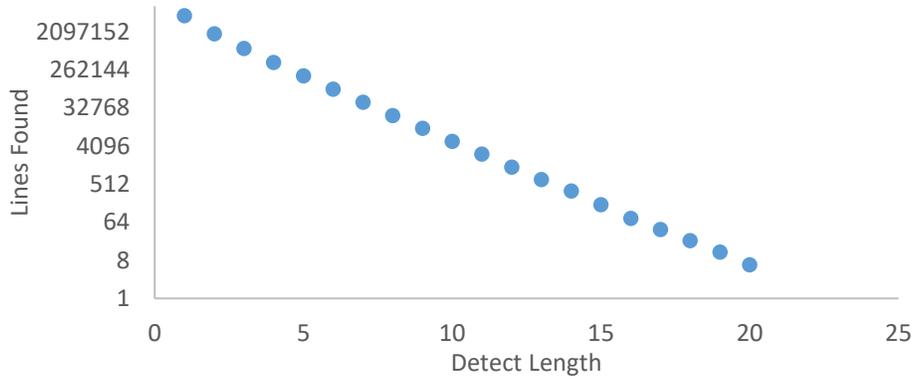

**Fig. 16.** Number of lines detected in $10^7$ random bits for various detect lengths.

**Table 1.** The lines found divided by the bitstream length and multiplied by the detect length.

| $n$ | $L_n$ |
|---|---|
| 1 | 0.499857 |
| 2 | 0.375009 |
| 3 | 0.24985 |
| 4 | 0.156031 |
| 5 | 0.093679 |

This table only shows the first 5 items in the series. By taking each item in $L_n$ and dividing it by the previous item, a new series called $L'_n$ is found as shown in Table 2. This series shows the percentage change of each item when compared to the previous. This series starts at $n = 2$ as there is nothing for $L_1$ to divide by. This series is notable as it can roughly be approximated as simple fractions, as shown in Table 3.

**Table 2.** $L'_n$ showing the percentage change of each item.

| $n$ | $L'_n$ |
|---|---|
| 2 | 0.750232 |
| 3 | 0.666251 |
| 4 | 0.624499 |
| 5 | 0.600385 |

**Table 3.** Data from Table 2 approximated as fractions.

| $n$ | Approx. $L'_n$ |
|---|---|
| 2 | 3/4 |
| 3 | 4/6 |
| 4 | 5/8 |
| 5 | 6/10 |

These fractions are observed to follow the form of $\frac{n+1}{2n}$. Using the percentage change of $\frac{n+1}{2n}$ between the values in $L_n$, the result shown in Table 4 is found. Note that $L_1$ is included despite not having an $L'_1$ to be multiplied by. This is because given a random bitstream of any size, the number of bits found for a detect length of 1 divided by the bitstream size is always $\frac{1}{2}$.

Table 4. $L_n$ as expressions for the percentage change between values.

| $n$ | $L_n$ |
|---|---|
| 1 | 0.5 |
| 2 | $L_1 \cdot \frac{n+1}{2n}$ |
| 3 | $L_2 \cdot \frac{n+1}{2n}$ |
| 4 | $L_3 \cdot \frac{n+1}{2n}$ |
| 5 | $L_4 \cdot \frac{n+1}{2n}$ |

It is noted that this new definition of $L_n$ defines every item as a multiple of all of the previous items. This when expanded is shown in equation 3.

$$L_n = 0.5 \cdot \frac{i+1}{2i} \cdot \frac{i+2}{2(i+1)} \cdot \frac{i+3}{2(i+2)} \cdot \ldots \cdot \frac{i+(n-1)}{2(i+(n-2))} \quad (3)$$

where $i$ is equal to the value of $n$ taken from its earliest use in the series, which in this case is $L_2$, but it will be left as $i$ for now. From this, all instances of $(i + x)$ in every numerator can be canceled using the next term's denominator. This means, however, that the first term retains its denominator, and the last term retains its numerator as visible in equations 4 and 5.

$$L_n = 0.5 \cdot \frac{i+1}{2i} \cdot \frac{i+2}{2(i+1)} \cdot \frac{i+3}{2(i+2)} \cdot \ldots \cdot \frac{i+(n-1)}{2(i+(n-2))} \quad (4)$$

$$L_n = 0.5 \cdot \frac{1}{2i} \cdot \frac{1}{2} \cdot \frac{1}{2} \cdot \ldots \cdot \frac{i+(n-1)}{2} = 0.5 \cdot \frac{i+(n-1)}{2^{n-1}i} \quad (5)$$

For the next step, the variable $i$ can be replaced with its known value of 2 to produce equation 6.

$$L_n = 0.5 \cdot \frac{2+(n-1)}{2^{n-1}(2)} = 0.5 \cdot \frac{n+1}{2^n} \quad (6)$$

Finally having a general form for $L_n$, an equation to get the probability of lines appearing for any detect length can be found by simply undoing the multiplication of detect length done to create $L_n$ in the first place, as shown in equation 7.

$$\frac{0.5 \cdot \frac{n+1}{2^n}}{n} = \frac{(0.5)(n+1)}{2^n n} = \frac{n+1}{(2^{n+1})n} \quad (7)$$

This yields the decimal probability of lines being found for any detect length of $n$, but needs to be multiplied by the bitstream length to determine exactly how many lines are found for any bitstream size. Multiplying the above equation by the bitstream length yields Equation 1.

## 5 Conclusion and Future Plans

A set of the data quality tests have been designed and tested to be used with the future data streams obtained from both setup designs [6, 7]. Additionally, the software that implements these tests is being produced and tested. It will implement additional commonly used methods for the testing of the randomness in the data as well.

Additionally, plans include completing the data collection of the large samples and testing them with the methods described in this work. This will be presented in the subsequent publications. We expect that the use of the FLS Test would be able to clearly show that the high/low method of processing is inferior to the even/odd method. Further objectives for the future of this project are to sync the MPPC + Arduino with an LED that is voltage controlled through means besides PWM. A proper enclosure for the board and sensor is also necessary, as well as a faster board to increase the number of amplitudes per second.

**Acknowledgements**: This work was supported by Clayton State University College of Arts & Sciences UCARE grant, FY 2023-24, and in part by NSF LEAPS-MPS Award 2316097.